\documentclass{article}
\usepackage{spconf,amsmath,graphicx}
\usepackage{bm}
\usepackage{multirow}
\usepackage{cite}
\usepackage{makecell}
\usepackage{hyperref}
\usepackage[hyphenbreaks]{breakurl}
\newcommand\toprule{\Xhline{.10em}}
\newcommand\midrule{\Xhline{.08em}}
\newcommand\bottomrule{\Xhline{.10em}}

\newcommand{\tabincell}[2]{\begin{tabular}{@{}#1@{}}#2\end{tabular}}

\title{IMPROVING SEQUENCE-TO-SEQUENCE VOICE CONVERSION BY ADDING TEXT-SUPERVISION}
\name{Jing-Xuan Zhang$^1$, Zhen-Hua Ling$^1$, Yuan Jiang$^2$, Li-Juan Liu$^2$, Chen Liang$^3$, Li-Rong Dai$^1$\thanks{This work was partially funded by the National
Nature Science Foundation of China (Grant No. 61871358) and the Key Science and Technology Project of Anhui Province under Grant No. 17030901005.}}
\address{$^1$National Engineering Laboratory for Speech and Language Information Processing,\\
University of Science and Technology of China, Hefei, P.R.China\\
$^2$iFLYTEK Research, iFLYTEK Co., Ltd., Hefei, P.R.China\\
$^3$Anhui Science and Technology Research Institute, Hefei, P.R.China\\
{\small \tt nosisi@mail.ustc.edu.cn, \{zhling,lrdai\}@ustc.edu.cn, }\\
{\small \tt \{yuanjiang,ljliu\}@iflytek.com, 279141996@qq.com}}

\begin{document}
\ninept
\maketitle
\begin{abstract}
This paper presents methods of making using of text supervision to improve the performance of  sequence-to-sequence (seq2seq) voice conversion.
Compared with conventional frame-to-frame voice conversion approaches, the seq2seq acoustic modeling method proposed in our previous work achieved higher naturalness and similarity.
In this paper, we further improve its performance by utilizing the text transcriptions of parallel training data. 
First, a multi-task learning structure is designed which adds auxiliary classifiers to the middle layers of the seq2seq model and predicts linguistic labels as a secondary task.
Second, a data-augmentation method is proposed which utilizes text alignment to produce extra parallel sequences for model training.
Experiments are conducted to evaluate our proposed method with training sets at different sizes.
Experimental results show that the multi-task learning with linguistic labels is effective at reducing the errors of seq2seq voice conversion.
The data-augmentation method can further improve the  performance of seq2seq voice conversion when only 50 or 100 training utterances are available.
\end{abstract}
\section{Introduction}
\label{sec:intro}

Voice conversion (VC) aims to convert the speech of a source speaker into that of target while keeping linguistic content unchanged \cite{
Childers1989Voice}. The VC technique has various applications such as identity switching in a text-to-speech (TTS) system, vocal restoration in cases of language impairment and entertainment applications \cite{
Mohammadi2017An}.

The most widely-used approach for voice conversion is adopting a statistical acoustic model to capture the relationship between
acoustic features 
of source and target speakers. In conventional method, frame-aligned training data is first prepared using dynamic time wrapping algorithm (DTW) \cite{2007Dynamic}. Then, an acoustic model is trained based on the paired source-target frames. During conversion, a mapping function is derived from the acoustic model, and target acoustic features are predicted from those of source frame by frame.
The acoustic model can be a joint density Gaussian mixture model (JD-GMM) \cite{Kain1998Spectral,
Toda2007Voice}, a deep neural network (DNN) \cite{
Desai2010Spectral,Chen2014Voice} or a recurrent neural network (RNN) \cite{Sun2015Voice,Lai2017Phone}.

Our previous work \cite{zhang2018} proposed a sequence-to-sequence (seq2seq) method 
for VC. A Seq2seq ConvErsion NeTwork (SCENT) is designed to model pairs of input and output acoustic feature sequences directly without explicit frame-to-frame alignment. The SCENT followed the encoder-decoder with attention architecture \cite{sutskever2014sequence,bahdanau2015neural,luong2015effective,wang2017tacotron,shen2017natural}.
This method achieved effective duration conversion, higher naturalness and similarity compared with conventional GMM and DNN-based methods \cite{zhang2018}. However, utterances converted by the seq2seq method may have mispronunciations and other instability problems such as repeating phonemes and skipped phonemes.

In practical voice conversion tasks with parallel training data, text transcriptions of both speakers 
are usually available. 
Thus, this paper presents methods of utilizing text-supervision to improve the seq2seq VC model.
First, a multi-task learning structure is designed. Auxiliary classifiers are added to the output layer of the encoder and the input layer of the decoder RNN,
 and are trained to predict the linguistic labels from the hidden vectors. 
Thus, the middle layers of the seq2seq model are regularized by the secondary task to be more linguistic-related, which is expected to reduce the issue of  mispronunciations at conversion time.
Second, a data-augmentation method is proposed by utilizing the text alignment information.
In previous seq2seq VC method, the whole utterances are used as the sequences for model training.
In order to increase the generalization ability of the trained seq2seq model, additional parallel fragments of utterances are derived using the alignment points given by text transcriptions, and are used as training samples.

The proposed method is evaluated using training sets of different sizes.
Experimental results show that our method of adding text supervision to seq2seq VC can generate utterances with higher naturalness and sometimes better similarity. The multi-task learning structure is effective at reducing pronunciation errors. The proposed data-augmentation method can further improve the model performance when the training set contains only 50 or 100 utterances.

\section{Previous Work}
\label{sec:previouswork}

\subsection{Related work}
\label{subsec:relatedwork}

Methods of incorporating text information in VC task have been investigated in previous studies. Text information was usually used as restrictions to improve the alignment between acoustic feature sequences \cite{tao2010supervisory, Lai2017Phone}. A CART-based voice conversion system was proposed in which phonetic information was used to grow the decision tree \cite{bonafonte2004including}. A phone-aware LSTM-RNN for VC was proposed \cite{Lai2017Phone}, which combined the monophones and spectral features as model inputs. Compared with previous studies, text transcriptions are utilized for improving the training of the seq2seq acoustic model and are not used at the conversion time in our proposed method. 

Multi-task learning has produced good results in various tasks, such as automatic speech recognition \cite{seltzer2013multi, chen2015speech}, speech synthesis \cite{wu2015deep} and natural language processing \cite{collobert2008unified}. In  DNN-based speech synthesis, a secondary task to predict a perceptual representation of target speech was proposed \cite{wu2015deep} to improve the perceived quality of generated speech.
Yang et al. \cite{yang2017statistical} proposed a generative adversarial networks (GAN)-based speech synthesis framework with phoneme classification. In our method, auxiliary classifiers for linguistic labels are added to the middle layers of the model. This secondary task provides additional supervision during training, which is expected to ``steer'' the hidden layers towards more linguistic-related representations. 




On image processing tasks, cropping images is common approach of data augmentation \cite{perez2017effectiveness}. 
In this paper, we propose to slice fragments from parallel utterances according to text alignment and use them as training samples. This technique could make use of more alignment information within the parallel utterances 
and is expected to reduce overfitting of the built seq2seq model.


\subsection{Sequence-to-sequence voice conversion}
\label{subsec:seq2seqvc}

\begin{figure}[t]
    \centering
    \centerline{\includegraphics[width=0.9\linewidth]{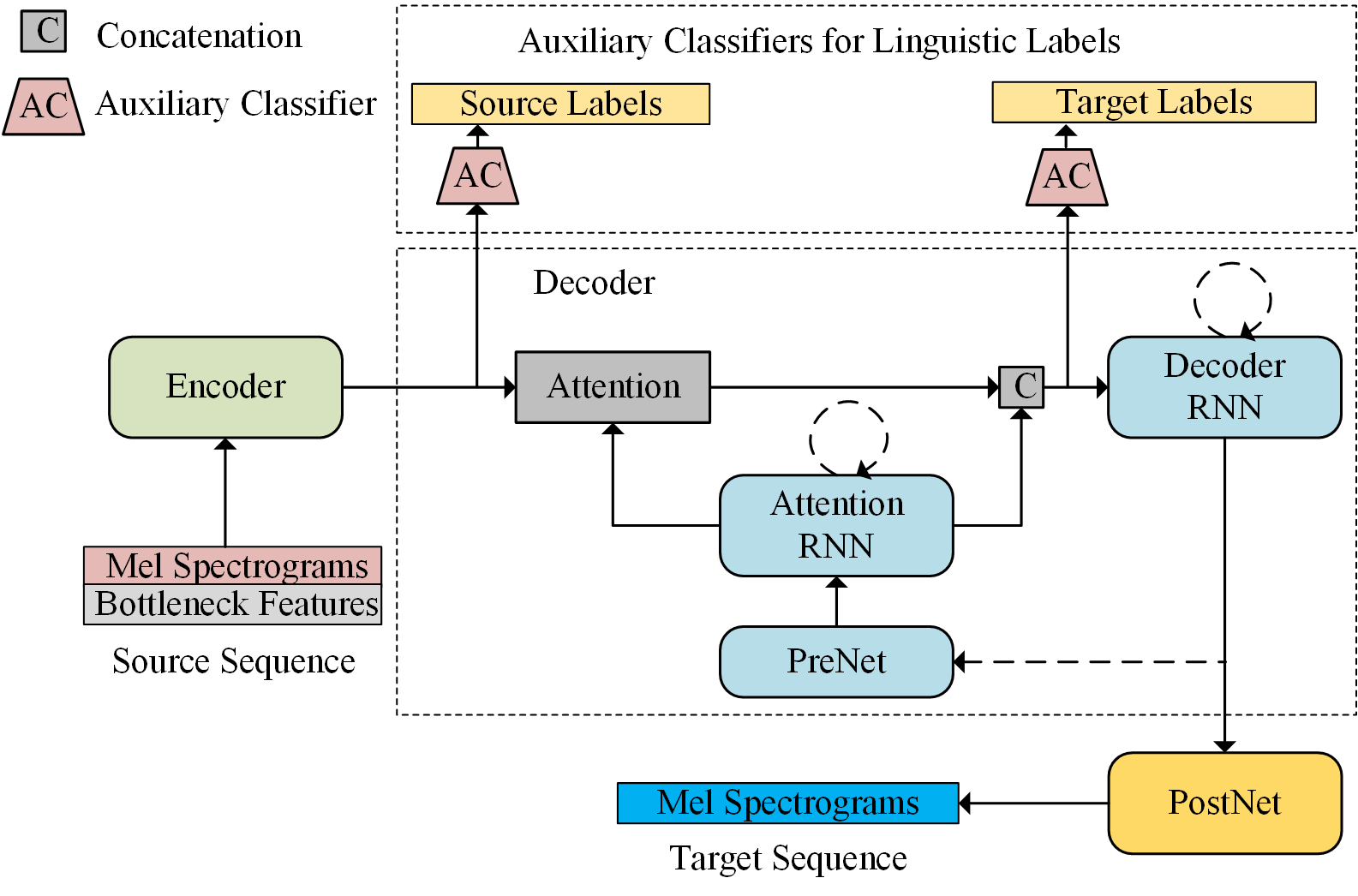}}
    \caption{Block diagram of Sequence-to-sequence ConvErsion NeTwork (SCENT) with auxiliary classifiers. The dashed lines represent recurrent or auto-regressive connections.}
\label{fig:digram}
\end{figure}

In our previous work \cite{zhang2018}, we proposed SCENT, a seq2seq acoustic model for VC.
Ignoring the component of auxiliary classifiers, \figurename~\ref{fig:digram} shows the structure diagram of SCENT, which follows the popular encoder-decoder with attention architecture.
Specifically, it is composed of an encoder, a decoder with attention and a post-filtering network (PostNet).

The input sequence of the model is the concatenation of mel-spectrograms and bottleneck features of source utterance. The bottleneck features are linguistic-related features which are extracted from speech signal using a speaker-independent automatic speech recognition (ASR) model. The encoder accepts input sequence and transforms it into hidden representations which are more suitable for the decoder to deal with.  At each decoder time step, the previous generated acoustic frame is fed back into a preprocessing network (PreNet), the output of which is passed through an attention RNN. The output of the attention RNN is processed by the attention module, which produces a summary of encoder output entries by weighted combination. The weighting factors are attention probabilities. Then the concatenation of this summary and the output of attention RNN is passed through the decoder RNN to predict output acoustic frame. In order to enhance the quality of the prediction, a PostNet is further employed 
to produce the final mel-spectrograms of target speaker. At last, a WaveNet neural vocoder \cite{denoord2016wavenet} conditioned on mel-spectrograms is utilized for the waveform reconstruction.


\section{Proposed Methods}
\label{sec:proposed}
Linguistic labels, such as phoneme identity, are firstly extracted from the text transcriptions and then aligned to source and target utterances respectively at the data preparation stage. 
The alignment can be obtained by manual annotation or automatic methods such as force alignment using a hidden Markov model (HMM).
Two methods of making use of the text supervision to improve the performance of the seq2seq VC model are introduced in this section.

\subsection{Multi-task learning with linguistic labels}
\label{subsec:auxiliary}



In parallel with learning to predict the acoustic features of target speaker, a secondary task is conducted to predicted linguistic labels from middle layers of the model. As presented in \figurename~\ref{fig:digram}, two auxiliary classifiers are added to the outputs of encoder and the inputs of decoder RNN. In each classifier, the input hidden representations are first passed through a dropout layer for increasing generalization. Then, the outputs of the dropout layer are projected to the category number of linguistic labels followed by a softmax operation. The targets of  the two classifiers are the linguistic labels that current hidden representations of encoder and decoder RNN correspond to respectively.
The cross-entropy losses of these two classifiers are weighted and added with the original loss of mel-spectrograms for training the model.

The auxiliary classifiers are designed for improving the seq2seq VC model by using stronger supervision from the text.
Intuitively, they help to guide the model to generate more meaningful intermediate representations which are linguistic-related. Adding classifier to both the encoder and decoder part is also supposed to help the attention module to predict correct alignments.
It should be noticed that the classifiers are only used at the training stage and are discarded at the conversion time. 
Therefore, no extra input and computation are required during conversion.

\begin{figure}[t]
    \centering
    \centerline{\includegraphics[width=0.75\linewidth]{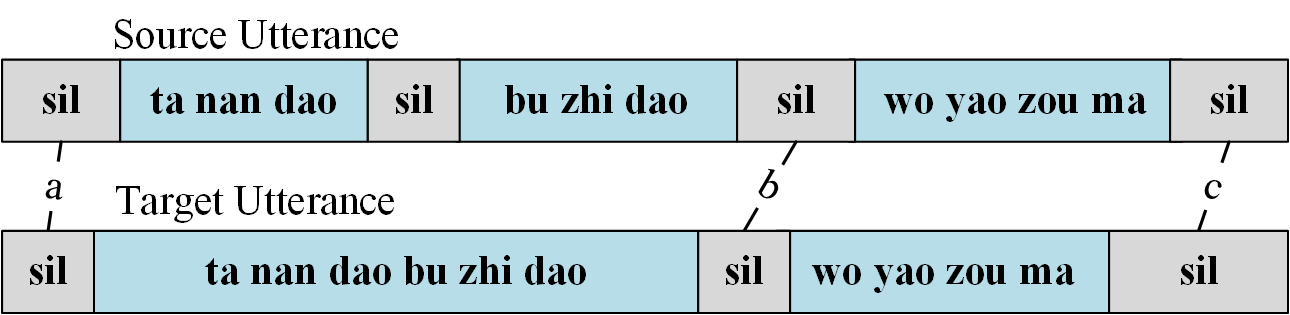}}
    \caption{Example of the aligned texts for a pair of parallel utterances. Texts are presented as Chinese pinyin and ``sil'' represents the silence.
    ``a'',``b'' and ``c'' represent totally three alignment points in this pair of parallel utterances. Therefore, totally $C_3^2$ (i.e., 3) potential parallel fragments can be sliced from original utterances.}
\label{fig:data}
\end{figure}

\subsection{Data-augmentation by text alignment}
\label{subsec:segmented}


\begin{table*}[t]
  \renewcommand\arraystretch{1.3}
\caption{MCDs and $F_0$ RMSEs on test set using training sets of different sizes. The minimum MCD and $F_0$ RMSE for each set size and speaker pair are highlighted in bold fonts.}\label{tab:ojective}
  \centering
  \resizebox{0.9\textwidth}{!}{
  \begin{tabular}{l|c|c|c|c|c|c|c|c|c|c|c|c}
  \toprule
  \multirow{4}*{\tabincell{c}{Size of \\training\\ data}}&
  \multicolumn{6}{c|}{Male-Female}&
  \multicolumn{6}{c}{Female-Male}\\
  \cline{2-13}
  &\multicolumn{2}{c|}{Seq2seq}&
  \multicolumn{2}{c|}{Seq2seq-MT}&
  \multicolumn{2}{c|}{Seq2seq-MT-DA}&
  \multicolumn{2}{c|}{Seq2seq}&
  \multicolumn{2}{c|}{Seq2seq-MT}&
  \multicolumn{2}{c}{Seq2seq-MT-DA}\\
  \cline{2-13}
  & MCD& $F_0$ RMSE& MCD& $F_0$ RMSE& MCD& $F_0$ RMSE & MCD& $F_0$ RMSE& MCD& $F_0$ RMSE& MCD& $F_0$ RMSE\\
  &(dB)&       (Hz)&(dB)&       (Hz)&(dB)&       (Hz) &(dB)&       (Hz)&(dB)&       (Hz)&(dB)&       (Hz)\\
  \midrule
  50 & 5.116 & 50.035 & 4.870 & 46.962 & \textbf{4.473} & \textbf{44.235} & 4.888& 56.496& 4.640& 54.023& \textbf{4.076} &\textbf{46.094} \\
  \hline
  100& 4.679 & 42.830 & 4.360 & \textbf{38.290} & \textbf{4.161} & 39.101 & 4.339 & 51.159& 4.022& 47.189&
  \textbf{3.970}& \textbf{43.621} \\
  \hline
  200& 4.130 & 40.976 & 4.000 & 35.583 & \textbf{3.936} & \textbf{35.558} & 3.920 & 45.733& \textbf{3.811}& \textbf{45.166}& 3.813 &45.515\\
  \hline
  400& 3.959 & 36.765 & 3.905 & 35.787 & \textbf{3.864} & \textbf{35.727} & 3.726 & 45.471& \textbf{3.696}&
  \textbf{44.952} &3.722 &45.401\\
  \hline
  1000& 3.802& 33.374 & 3.776 & \textbf{32.155} &\textbf{3.774} & 34.604  &3.556 & \textbf{41.748}& 3.579 &
  43.652 &\textbf{3.545} &44.107\\
  \bottomrule
  \end{tabular}}
\end{table*}


In our previous seq2seq VC method, pairs of whole utterances are used as the input and output sequences for model training. 
With text alignments, intra-utterance alignments can also be utilized to produce more sequence pairs.

In our method, an ``alignment point'' is defined as a common silence fragment in a pair of parallel utterances. \figurename~\ref{fig:data} presents an example for illustration.
Parallel fragments, which contain the same linguistic contents within two utterances, are extracted by  selecting two alignment points as the starting and ending positions. The reason that alignment points are defined at silences is to make sure that the parallel fragments are less influenced by surrounding contents.
For a pair of parallel utterances containing $N$ alignment points, totally $C_N^2$  parallel fragments can be extracted.
When processing each pair of utterances at training time, a pair of parallel fragments are randomly selected  from all $C_N^2$  possibilities instead of using
the whole utterances.





\section{Experiments}
\label{sec:experiments}

\subsection{Experimental conditions}
\label{subsec:expconditions}

Our dataset for experiments contained 1060 parallel Mandarin utterances of one male speaker (about 53 min) and one female (about 72 min) speaker, which were separated into a training set with 1000 utterances, a validation set with 30 utterances and a test set with 30 utterances. Smaller training sets containing 50, 100, 200 and 400 utterances were also constructed by randomly selecting a subset of the 1000 utterances for training. The recordings were sampled at 16kHz. 80-dimensional mel-scale spectrograms were extracted every 10 ms with Hann windowing of 50 ms frame length and 1024-point Fourier transform. 512-dimensional bottleneck features were extracted using an ASR model every 40 ms and were then upsampled by repeating to match the frame rate of mel-spectrograms. Text transcriptions were firstly converted into sequences of phonemes with tone using a rule-based grapheme-to-phoneme model. The phoneme with tone sequences were then aligned to the speech using an HMM aligner.

Details of the seq2seq model and the WaveNet vocoder were kept the same as our previous work \cite{zhang2018}. The output layer of the decoder in SCENT was a mixture density network (MDN) 
layer with 2 mixture components.

We used the batch size of 4 and Adam optimizer \cite{kingma2014} for model training. The learning rate was 0.001 in the first 20 epochs and exponentially decay 0.95 for 50 more epochs. 
For WaveNet training, the $\mu$-law companded waveforms were quantized into 10 bits. The learning rate was $10^{-4}$.
The focus of this paper was acoustic modeling, not WaveNet vocoder.
Therefore, the WaveNet vocoder of each speaker was trained using the waveforms of his or her full training set for convenience.

Three methods were compared in our experiments. The configuration of each method is described as follows~\footnote{Audio samples are available at \url{https://jxzhanggg.github.io/Text-supervised-Seq2SeqVC}.}:

\begin{itemize}
\item \textbf{Seq2seq:} Baseline method using previous proposed sequence-to-sequence acoustic model \cite{zhang2018}.
\item \textbf{Seq2seq-MT:} 
    Improving the baseline method using the multi-task learning structure proposed in Section~\ref{subsec:auxiliary}. Auxiliary classifiers were adopted for predicting linguistic labels at training time. Each classifier contained two separated linear projection with the softmax activation for predicting phoneme identity and tone category simultaneously. The weighting factors for phoneme and tone classification were  0.1 and 0.05 respectively, which were tuned on the validation set.
\item \textbf{Seq2seq-MT-DA:} In addition to multi-task learning, the data-augmentation method introduced in Section~\ref{subsec:segmented} was also adopted. In our full training set, the average number of alignment points in each pair of utterances was 3.15. The learning rate was fixed in first 40 epochs for better model convergence.
    We also tried to use larger batch size because the average length of each training sample became shorter. However, the results showed no improvement on the validation set.
\end{itemize}

\subsection{Objective evaluation}
\label{subsec:objective}

$F_0$ and mel-cepstra were extracted from the converted utterances using STRAIGHT \cite{Kawahara1999Restructuring}. Then, mel-cepstrum distortions (MCD) and root mean square error of $F_0$ ($F_0$ RMSE) on test set were reported  in  \tablename~\ref{tab:ojective}. From the table, we can see that all methods obtained lower MCD and $F_0$ RMSE given more training data. 
When the training data was limited, i.e. only 50 or 100 training utterances available, the proposed method using multi-task learning outperformed the baseline seq2seq method with a large margin. Adopting the data-augmentation method can further improve the performance of acoustic models. 
When more training data became available (e.g., 200 and 400 utterances), the performances of the Seq2seq-MT and Seq2seq-MT-DA methods were close and still better than the baseline method. When training with all parallel data, the proposed method obtained close MCD but higher $F_0$ RMSE than the baseline method. 

In summary, the proposed method achieved lower objective error when the training set contains 50, 100, 200 and 400 utterances respectively.
Compared with Seq2seq-MT, the Seq2seq-MT-DA method can further improve the prediction accuracy  when the size of training data was 50 and 100. 
When training with 1000 utterances, no significant objective improvement  was observed after data augmentation.
The reason may be that the fragments used after data augmentation neglected the influence of their contexts in utterances. 
This negative effect may counteract the positive effect of reducing overfitting.
Besides, the MCD and $F_0$ RMSE of our proposed method was not better than the baseline method when models were trained with 1000 utterances. Subjective evaluations were  conducted to further investigate the effectiveness of our proposed method.

\subsection{Subjective evaluation}
\label{subsec:subjective}

The first subjective evaluation was conducted to evaluate  the stability of 
Seq2seq and Seq2seq-MT methods. A native listener was asked to identify the mistakes occurred in the test utterances converted using these two methods, which included mispronunciation, repeating phoneme, skipped phoneme and unclear voice. 
The counted numbers of mistakes are presented in \tablename~\ref{tab:submistake}.
The evaluation results indicate that 
multi-task learning with linguistic labels can alleviate the problem of instability under all size of training data.
A closer inspection on the mistakes of the Seq2seq method found that the main problem of instability was mispronunciation when the size of training data was relatively large, i.e. 400 or 1000 utterances. Converted utterances sometimes suffered from unnatural tone or incorrect phoneme.
When the size of training data got smaller, mistakes of skipped phone, repeating phone increased, which were usually caused by improper attention alignments. The multi-task learning could help to alleviate both kind of problems.

\begin{table}[t]
  \centering
   \renewcommand\arraystretch{1.3}
  \caption{Numbers of mistakes identified subjectively on test set under different sizes of training data and speaker pairs.}\label{tab:submistake}
  \resizebox{0.85\columnwidth}{!}{
  \begin{tabular}{l|c|c|c|c}
  \toprule
  \multirow{2}*{\tabincell{c}{Size of \\training data}}&
  \multicolumn{2}{c|}{Male-Female}&
  \multicolumn{2}{c}{Female-Male}\\
  \cline{2-5}
  & Seq2seq & Seq2seq-MT &Seq2seq &Seq2seq-MT\\
  \midrule
  50 & 43 & \textbf{27} & 38 & \textbf{30} \\
  \hline
  100& 22 & \textbf{16} & 23 & \textbf{15} \\
  \hline
  200& 19 & \textbf{12} & 16 & \textbf{6} \\
  \hline
  400& 16 & \textbf{9} & 10 & \textbf{6} \\
  \hline
  1000& 12& \textbf{3} & 7 & \textbf{3} \\
  \bottomrule
  \end{tabular}}
\end{table}

Furthermore,  ABX preference tests were conducted on both similarity and naturalness. Two conditions with 50 and 1000 training utterances were investigated. As we described in Section~\ref{subsec:objective}, when the size of training data was small, data augmentation method further improved the objective performance of the model. When training with 1000 utterances, no significant objective improvement  was observed after data augmentation.
Therefore, we compared Seq2seq  with Seq2seq-MT-DA for using 50 training utterances and Seq2seq with Seq2seq-MT for using 1000 training utterances respectively.
10 native listeners were involved in the evaluation. 20 sentences in the test set were randomly selected. The conversion results were presented for listeners in random order.

\begin{figure}[t]
  \centering
  \includegraphics[width=0.77\columnwidth]{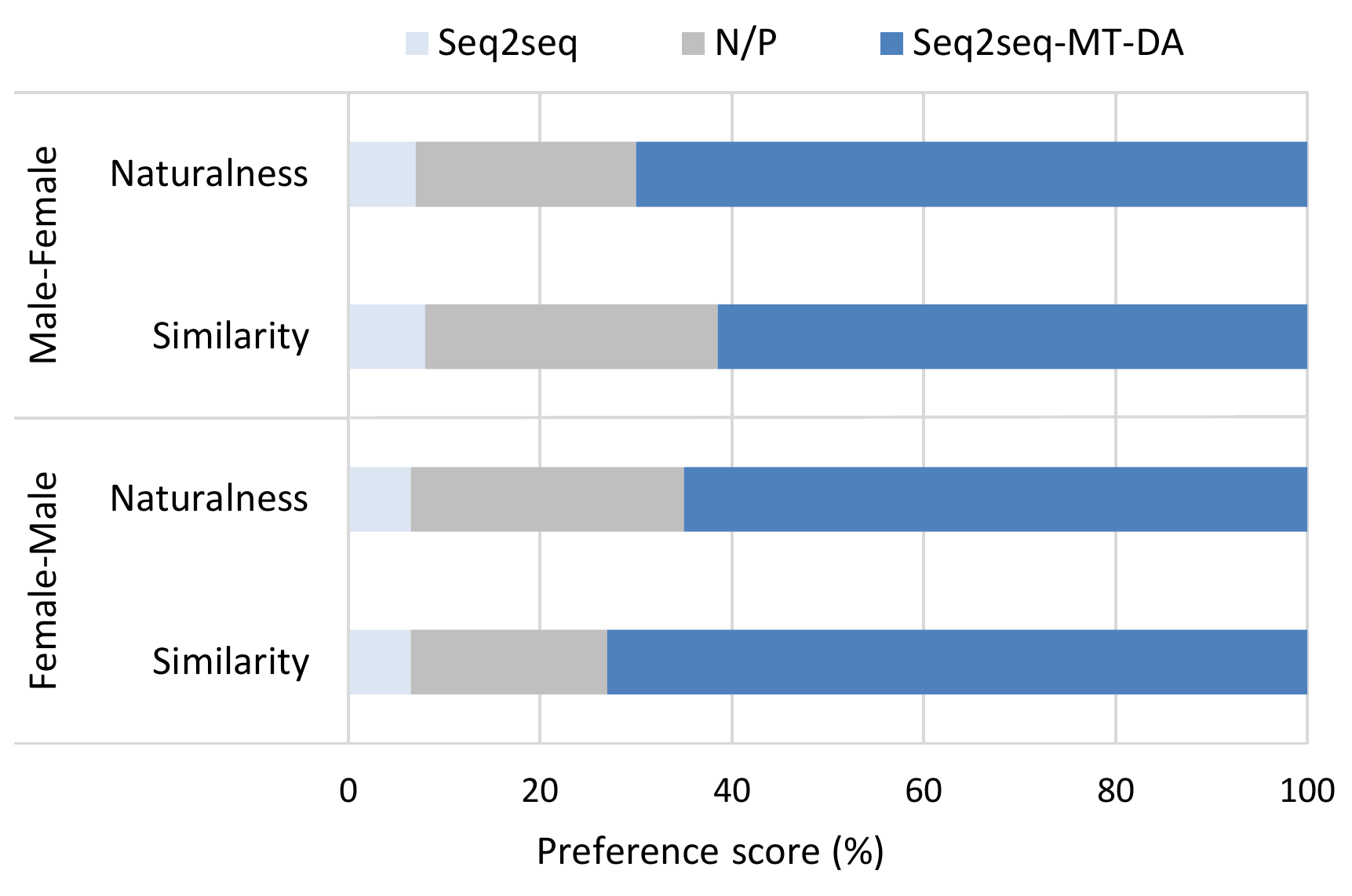}\\
  \caption{Results of ABX preference test for comparing Seq2seq and Seq2seq-MT-DA method under the condition of 50 training utterances.
  ``N/P'' represents no preference. The $p$-values of student $t$-test on the evaluation results  are $3.84\times10^{-33}$, $1.02\times10^{-24}$, $5.85\times10^{-37}$ and $5.83\times10^{-30}$ respectively.}\label{fig:AB50}
\end{figure}

\begin{figure}[t]
  \centering
  \includegraphics[width=0.77\columnwidth]{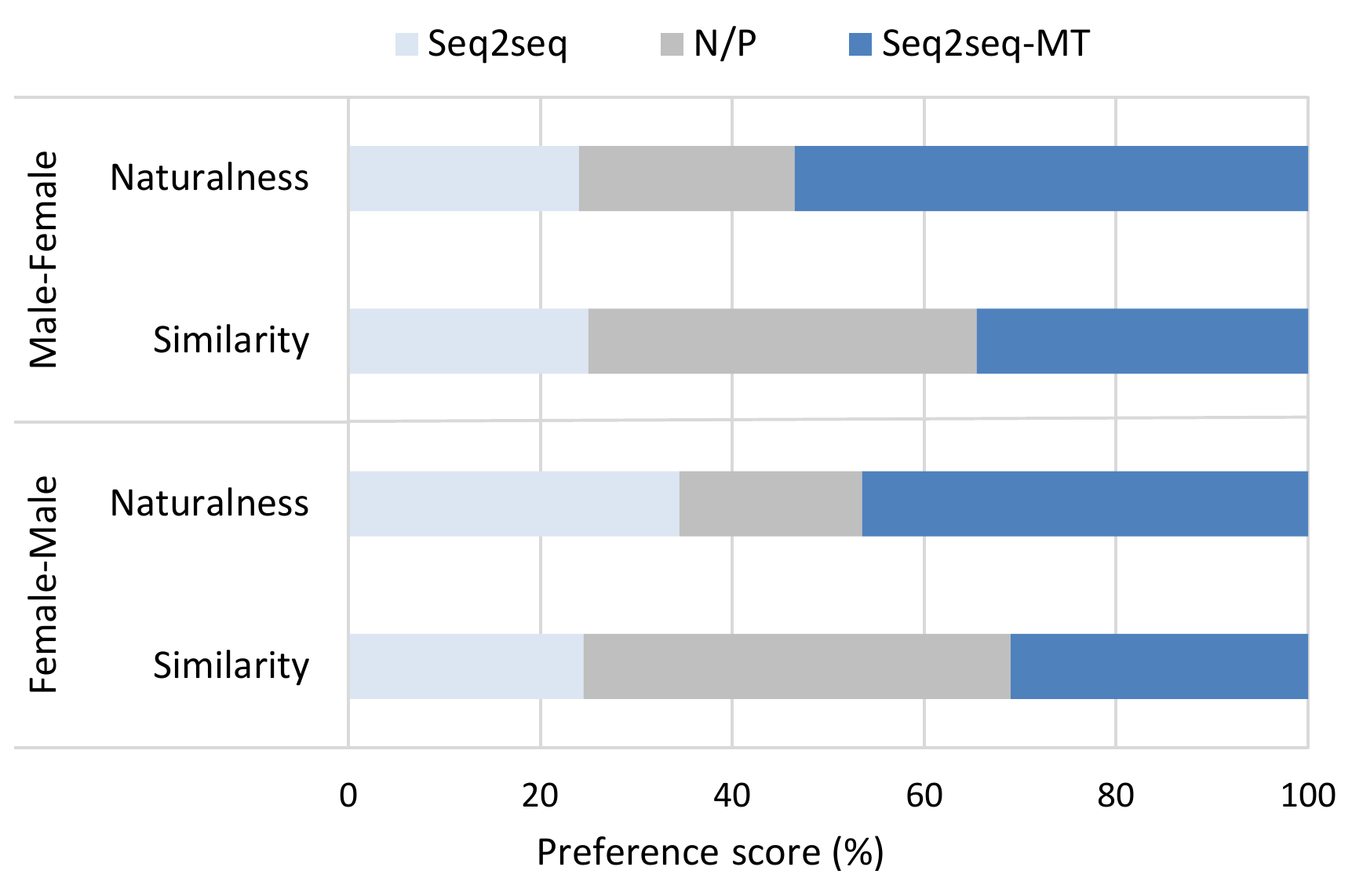}\\
  \caption{Results of ABX preference test for comparing Seq2seq and Seq2seq-MT method under the condition of 1000 training utterances. ``N/P'' represents no preference. The $p$-values of student $t$-test on evaluation results  are $1.16\times10^{-6}$, 0.0815, 0.0592 and 0.218 respectively.}\label{fig:AB1000}
\end{figure}

The experimental results are presented in \figurename~\ref{fig:AB50} and \figurename~\ref{fig:AB1000}. The evaluation results from \figurename~\ref{fig:AB50} show that the proposed Seq2seq-MT-DA method obtained significant higher preference score on both similarity and naturalness, which was consistent with the results of objective evaluations. These results indicate that the proposed methods improved model training significantly when the training data was limited.
\figurename~\ref{fig:AB1000} shows that the multi-task learning method improved the naturalness of converted speech when 1000 training utterances were available. The similarity improvement on female-to-male conversion was insignificant since the $p$-value was 0.218. 

\section{Conclusion}
This paper has presented two methods to improving seq2seq voice conversion by utilizing text supervision. 
First, a secondary task is introduced based on the framework of multi-task learning. Auxiliary classifiers are added for predicting corresponding linguistic labels from the middle layers of the model.
Second, a data-augmentation method is proposed, in which fragments of original utterances are randomly extracted at each training step. 
Experimental results validated the effectiveness of our proposed method for improving model training. The multi-task learning alleviates the instability problems, such as mispronunciations, in the conversion results of seq2seq model. The data-augmentation method can further improve the performance of seq2seq VC model with limited training data.

Although the proposed methods can enhance the seq2seq VC model effectively, the degradation of model performance is still significant when only small training sets are available. Future work includes  further improving the seq2seq model using other techniques in the resource-limited situation, such as model adaptation.


\small
\bibliographystyle{IEEEbib}
\bibliography{strings,refs}

\end{document}